# GUIDELINES FOR THE CREATION OF ANALYSIS READY DATA

Harriette Phillips, Aiden Price, Owen Forbes, Claire Boulange, Kerrie Mengersen, Marketa Reeves, Rebecca Glauert

Globally, there is an increased need for guidelines to produce high-quality data outputs for analysis. No framework currently exists that provides guidelines for a comprehensive approach to producing analysis ready data (ARD). Through critically reviewing and summarising current literature, this paper proposes such guidelines for the creation of ARD. The guidelines proposed in this paper inform ten steps in the generation of ARD: ethics, project documentation, data governance, data management, data storage, data discovery and collection, data cleaning, quality assurance, metadata, and data dictionary. These steps are illustrated through a substantive case study that aimed to create ARD for a digital spatial platform: the Australian Child and Youth Wellbeing Atlas (ACYWA).

## 1. INTRODUCTION

In almost every field of research there is an increased need to process quality-assured data for analysis [1], [2]. The production of analysis ready data (ARD) should adhere to any lawful and ethical requirement and be both interoperable with other datasets and throughout time [1], [3], [4], [5], [6]. ARD should also be created through minimal effort and allow for immediate analysis [1]. In response to the need for ARD, various criteria have emerged over the past five years [7], [8], [9]. These works typically focus on a specific subset of components of ARD development, without consideration of a holistic set of process guidelines, causing issues in consistency and comparability [9], [10]. In this section, key articles in the literature are critically reviewed and summarised.

An article published in 2017 produced guidelines for three ARD steps: data discovery, data collection, and data cleaning [9]. These steps enable data to be collected and compiled into a standardised format before its final release. A later paper, published in 2018, largely focused on data discovery, data collection, and data storage addresses the ingestion and management of large volumes of data for social media analytics [11]. Ethics was also identified as an important issue in the production of ARD but was determined to be out of scope. In the following year, a book on these guidelines was published for managing and sharing data [8]. This comprehensive resource includes extensive information on data collection, data cleaning, data governance, data management, data storage, ethics, and metadata. It was also outlined that ethics and legal requirements are dependent on the region in which the research is being conducted. Building on this work, a 2021 contribution focuses on the topics of metadata and data dictionary with the aim of creating sharable and transparent data that users can interpret easily and preferably be both human-readable and machine-readable [12]. This work can be further extrapolated through the principles that are aimed at making data findable, accessible, interoperable, and reusable [13]. An article published in the same year focused on building ARD through creating accessible data through open data products [2]. A recent article further extended the range of steps in the production of ARD, namely on the topics of general data collection and quality assurance [10]. They aimed to improve data quality by discussing quality assurance, data collection and ethics. The Committee on Earth Observation Satellites has also developed a framework for the development of ARD specifically for the use of satellite data [14]. This framework mainly focuses on allowing for immediate analysis of ARD through a minimum set of specifications whilst following the core requirements for metadata.

In summary, the literature to date suggests ten steps that should be considered in the creation of ARD: ethics, data governance, project documentation, data management, data storage, data discovery and collection, data cleaning, quality assurance, metadata, and data dictionary. Current literature does not consider these steps holistically to provide a comprehensive approach in producing ARD. This paper aims to develop ten key steps, summarised from current literature, which can then be used to create high-quality ARD outputs. The proposed guidelines will be investigated in Section 2 and contextualised, through a digital atlas, in Section 3.

Digital atlases are typically online, interactive tools that facilitate the display, inspection, and comparison of geospatial and temporal data [15]. The specific exemplar, used as a case study to develop the proposed guidelines, was the Australian Child and Youth Wellbeing Atlas (ACYWA) [16]. This digital atlas aims to improve the health and wellbeing metrics for Australian children and young people. Consequently, the production of high-quality ARD was essential for the production and interpretation of this digital atlas.

## 2. GUIDELINES FOR THE CREATION OF ANALYSIS READY DATA

The proposed guidelines have been compiled into a comprehensive approach for processing, organising, developing, and producing ARD. Each principle is discussed in detail in the subsequent subsections: 2.1 Ethics, 2.2 Project Documentation, 2.3 Data Governance, 2.4 Data Management, 2.5 Data Storage, 2.6 Data Discovery and Collection, 2.7 Data Cleaning, 2.8 Quality Assurance, 2.9 Metadata, and 2.10 Data Dictionary.

The ten principles that form the proposed guidelines for producing ARD are summarised in Figure 1 and Table 1. The sequential ordering was determined by the order necessary for the case study (ACYWA). The process typically starts with an ethics application, establishing project documentation, and leads into the other generation steps to create three outputs:

metadata, data dictionary, and analysis ready data. Establishment of data storage is also required before data can be accessed. Thus, data storage needs to occur before data can be collected. Generation steps often occur concurrently with one another (e.g., data cleaning and quality assurance). These steps can also be repeated and updated as required. This is represented by the dotted line shown in Figure 1.

**Figure 1.**

Visualisation of the Ten Guidelines for the Generation of Analysis Ready Data

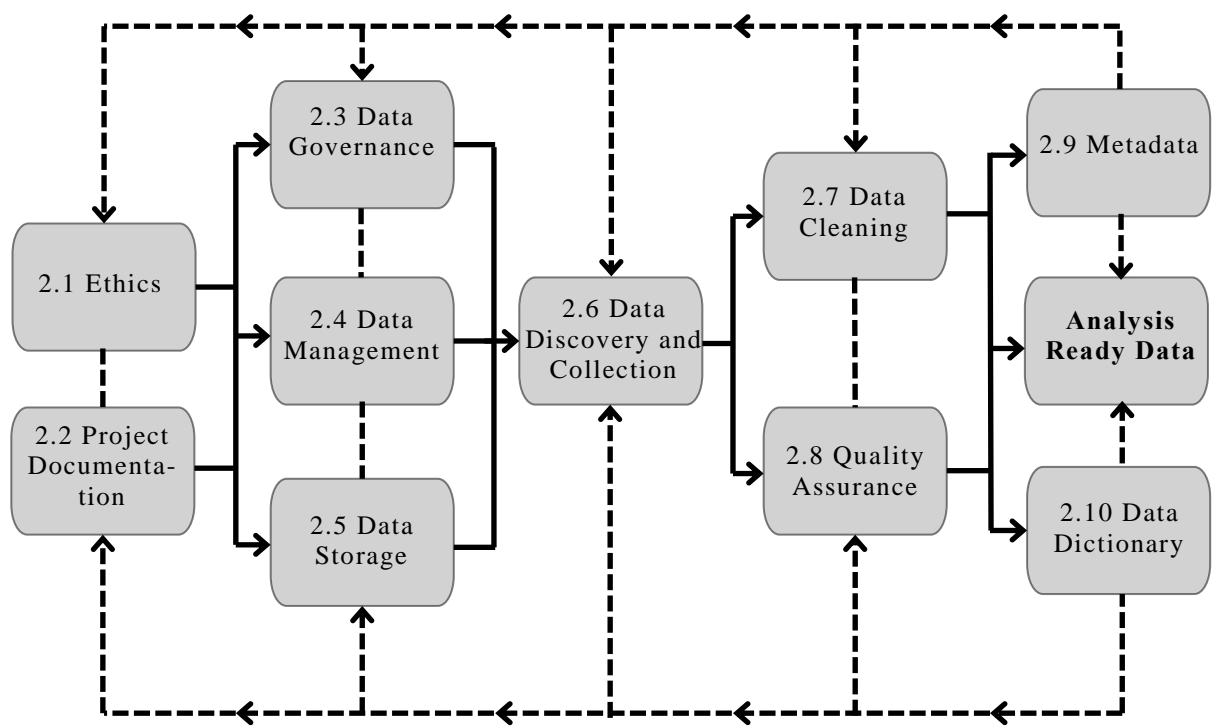

**Table 1.**

Summary of the principles for the guidelines in the production of high-quality ARD

| Principle | Consideration |
|---|---|
| Ethics | Ethics is typically established at the onset of a project where approval is sought for how methodologies will be carried out. This |
| | also governs other aspects of the project including, risks, participants, and research aims and outcomes, etc. These legal and ethical requirements are region and context specific. |
| Project Documentation | Project documentation includes the research aims, decisions, and outputs for a given project which should be established at the onset of a project and maintained throughout the entirety of a project's progress. |
| Data Governance | Data governance focuses on the protection of information surrounding data use through compliance with strategies to safeguard both person and non-personal information from misuse and mishandling. This includes compliance with the safe use of data sharing and collecting. |
| Data Management | The management of data and all steps related to the creation of ARD (e.g., data storage, data collection, data cleaning, quality assurance, metadata, and data dictionary). |
| Data Storage | Safe storage of materials mainly related to the data, where backup, retention, and deletion strategies should be implemented. |
| Data Discovery and Collection | Data discovery is the initial phase of data in alignment with project's methodologies and aims. Data is then collected through appropriate sources. These processes are well documented and should be guided by a data management plan. |
| Data Cleaning | Data cleaning occurs to compile data into a standardised format that does not contain identifiable information and is free of errors, inconsistencies, uncertainty, noise, etc. |
| Quality Assurance | Quality assurance is a process that occurs concurrently with other data processing steps to detect and resolve anomalies in data. This process is used to confirm that all processes have been consistently applied during the data processing steps. |
| Metadata | Metadata refers to the creation of supplementary information about dataset characteristics, preferably in human and machine-readable format. This information is collated using both automatic and manual processes. This information is used to complement the data to increase its visibility and reusability. |
| Data Dictionary | Data dictionaries are created to also provide supplementary information, ease of understanding and interpretation according to commonly used terms or data vocabulary within a given field. As opposed to metadata, this information is often supplied on an individual data indicator level to provide context about data processes, data sources, and any other necessary contextual information required for any further data analytics. |

## 2.1 Ethics

Ethics should be at the forefront of project development. Projects are obligated to conduct work in alignment with regional-specific legal and ethical requirements [1]. This is showcased through gaining ethical approval or clearance through an Ethics committee. Ethics often governs the entire process of project development (e.g., methodologies, data, participants, etc) where the approved ethics application typically outlines how the research should and will be conducted. The legal and ethical requirements should be researched before the commencement of a research project and abided by throughout the entire research process [8], [17]. Typically, ethics applications require review from a committee and adhere to an extensive list of criteria [18]. This is to ensure that the scope is achievable, often striving for improvements in a given field, whilst safeguarding participants. These criteria can be used for ARD creation. This provides projects with the necessary resources to develop a project in alignment with the region-specific legal and ethical requirements. The region-specific criteria also vary depending on the field of research (e.g., human research). This ensures that the ARD is generated based field of research a project is being conducted. Thus, ensuring that specific legal and ethical requirements of each criterion can be adhered to effectively and ethics can be subsequently for approved for individual projects. Examples of region and field-specific criteria for the creation of an ethics application include The Belmont Report on the Ethical Principles and Guidelines for the Protection of Human Subjects of Research (Belmont Report) and The National Statement on Ethical Conduct in Human Research (National Statement). These criteria have been detailed below.

The American standard, The Belmont Report, focuses on three main topics for conducting ethical research involving human participants: respect for persons, beneficence, and justice. Respect for persons details that participants are able to make informed decisions about their involvement in the research process and their own best interests [19]. This is understood through two ethical convictions. The first, that participants are autonomous and are entitled to be protected. Second, any participants with diminished autonomy are also entitled to be protected. Beneficence is aimed at minimising any harm and risks whilst maximising the benefits for participants who participated in the research. This is aimed at securing the well-being of participants through acts of kindness and charity that go beyond strict obligation. The last topic, justice, is that participants should be respected and treated equally.

Similarly, an Australian standard, the National Statement, can be used to develop, design, review, and conduct ethical human research. The guidelines in this approach require a review of values, judgement, principles, and context, depending on the level of risk to participants [20]. There are six topics in the national statement: merit and integrity in research, justice, beneficence, respect, risks and benefits, and general requirements for consent which are documented in Appendix 1. These topics provide a flexible framework for the design and review of ethical research.

**2.2 Project Documentation**

Projects need explanatory or contextual information surrounding data, procedures, decisions, and decision making processes to effectively conduct research [8]. The information surrounding data and processes is often referred to as project documentation [8], [21]. Project documentation is most easily maintained throughout the entirety of the research project when established at the outset of a project [8]. Furthermore, the inclusion of comprehensive documentation on how and why data were collected is critical for the short-, medium-, and long-term longevity of projects. Project documentation needs to be sufficient enough to ensure that data and project related materials are understandable, usable, and accessible to all users. For example, a data management plan (DMP) can be used to document the key information [22], [23]. This can include general project documentation such as documentation of decision making. Using a DMP to guide project documentation established a clear outline of project protocols. DMPs have been explored in further detail in Section 2.4 and Appendix 3. This documentation should also support effective strategies for maintaining the security of project information [21]. Effective security measures should include implementing version control when deleting and modifying records. These strategies should also include destruction methods after the project has been completed.

**2.3 Data Governance**

Data governance is achieved through the authoritative control and compliance with privacy, security, quality and ongoing data management in accordance with region-specific regulations [24], [25], [26]. These regulations aim to minimise risk and cost involved in data use whilst maximising the value of the data; and how it is used, shared, and collected [24]. Data governance is adopted to safeguard information minimising privacy concerns by preventing data from being misused or mishandled [25], [27]. Data governance is important for safeguarding both personal and non-personal information. Regulations differ depending on the type of data and the region in which the research is being conducted in [8], [25]. These regulations can be used to govern the quality of ARD. The regulations ensure that ARD are created and maintained at a high standard. Examples of these regulations include the General Data Protection Regulation (GDPR) and the Australian Code for the Responsible Conduct for Research (the Code) which are detailed below.

The GDPR is a European standard for regulating data governance strategies [28]. The GDPR was introduced to minimise the risk and cost involved in the collection, use, and sharing of personal data. This regulation provided a set of principles for data compliance and data protection, including data processing, lawfulness, consent rights, responsibilities, and documentation of activities [24], [26], [28]. The introduction of these stricter guidelines resulted in revisions by European researchers to how data is currently collected, shared, and used [25], [27], [28]. However, whilst the GDPR does provide great benefits and protections for personal data, the cost of compliance cannot be met at times due to insufficient resources [29], [30]. The GDPR is also a region-specific regulation, where there is the possibility of contextualising this approach to various regions, but compliance would not be required.

An alternative approach for managing the risk and cost relevant to data use is the Code. In contrast to the GDPR, this framework provides principles for the responsible conduct of research but does not incorporate laws, regulations, or guidelines [31]. The Code is reliant on the integrity of researchers and the expectation that principles and responsibilities for conducting honest and ethical research will be followed. However, failure to comply with the principles outlined in the Code can result in a breach or research misconduct. There are eight key principles outlined in the Code: honesty, rigour, transparency, fairness, respect, recognition, accountability, and promotion that are summarised in Appendix 2. There is limited current literature into the use of the Code, and it does not provide a comprehensive guideline for all aspects of data use, but it does use supporting documentation to provide researchers with expectations for their research [31], [32]. A recent study found that the principles contained within the Code were strongly endorsed [32].

## 2.4 Data Management

Effective management of large volumes of data is critical for individuals and organisations [22]. A Data Management Plan (DMP) is an effective way to summarise the key processes in the production of data [22], [23]. DMPs are often a research output requirement and having a well-managed and effective DMP can have a long-term term impact on the sustainability of a project. A detailed DMP enables the strategic implementation of good practices for generating data [8]. They provide documentation on the actual or expected aims so that they can be implemented with greater clarity of expectations and tasks. If a DMP is submitted to an ethics committee and approved this can validate that the project's aims will be ethically and legally sound and can be carried out for the duration of the project. Furthermore, developing a DMP can generate avenues for discussing issues, milestones, and progress. This will save time in the projection of ARD by reducing errors; duplication of efforts; assisting in the production of sharable data; improving data protection, security, quality; and promoting compliance with legal and ethical standards [22], [23]. A DMP generally involves key processes for ARD generation that should be documented and considered before the commencement of research. Appendix 3 summarises some considerations for the creation of an effective DMP based on current literature. This includes considerations for data discovery, data description, data collection, data organisation, and data preservation procedures, data storage, data volume, data loss procedures, data privacy and confidentiality, data ownership, quality assurance, documentation, dissemination, and metadata. It is also recommended that once a DMP has been established it should be updated at least annually and at the commencement of research. The DMP outputs should be easily shareable with other researchers [33].

## 2.5 Data Storage

Implementation of effective storage plans and backup strategies can prevent the need to replicate research, minimise reputational risk, and maintain the privacy of research subjects [8]. Supporting documentation for The Code (see Section 2.3) outlines key storage, retention, and disposal strategies and should occur regardless of whether the project is occurring internally or externally. These policies include that projects should comply with current licensing and copyright requirements, and be conducted in accordance with research specific practices and standards [34]. The research should also be conducted in compliance with any ethical; privacy; and publication requirements, as well as compliance with laws; guidelines; and regulations. Furthermore, data storage should also be considered when completing a DMP [22], [23]. A DMP considers a wide range of items that capture how data should be stored. This ensures that, from the onset of a project, the data storage for a given project is apparent. The full list of considerations for a DMP has been detailed in Appendix 3.

Effective data storage, including avoiding data loss and maintaining data retention, is necessary beyond the immediate scope of a project [8]. This is crucial for a project's long-term success. Cloud-based storage is most commonly used with advanced technology and when large amounts of data are ingested. As such, it has recently gained popularity for use by both researchers and end users where data, documentation and information related to the research can be reliably accessed from any device, and large-scale data ingest is supported [35]. Cloud-based storage also provides a flexible solution for cross-institutional access, is cost-effective, and provides larger storage capacities than other alternatives.

## 2.6 Data Discovery and Collection

Discovering and sourcing appropriate data should be achieved in alignment with project methodologies and aims. This step should provide adequate justification for the inclusion of given data before it is collected [11]. For example, this justification can include alignment of data with methodologies and aims by identifying appropriate sources with a standardised set of data characteristics (e.g., when and where the data were collected, spatial and temporal granularity, characteristics of participants). This discovery process is key in any given research to describe, measure or reflect the condition of interest through the use of data [36]. Data can then be collected through the appropriate sources with reference to the pre-determined set of data characteristics. This process should be well documented and considered in the initial phases of creating, organising, and managing data that is continued through the entire research process [8]. Data collection and discovery is typically guided by a clear DMP, where the data characteristics, expected volume of data, start date, and end date of data collection should be documented when compiling a DMP (see section 2.4 Data Management and Appendix 3) [36]. The inclusion of data discovery and collection in a DMP allows for the roles, responsibilities, and expectations of data collectors to be clearly defined. This ensures that all parts of the data discovery and cleaning process have been considered, which can include items such as the expected volume of data, expected data characteristics, start date of collection, and end date of collection [33].

Data are produced in different formats of varying quality [2], [37]. These data can be produced in several formats such as numerical, text, or multimedia. The varying degree of quality can be a result of noise, inconsistencies, and incompleteness in the data which can limit its use [2], [38]. This can also be referred to as uncertainty, where the absence of lack or absence of information, created by noise, inconsistencies and incompleteness, in data creates uncertainty during interpretation (see Section 2.8) [38], [39]. Subsequently, this varying degree of quality and formatting as well as the presence of uncertainty means data is often not readily usable in its original form [2]. Processes are required after collection occurs to ensure that data is in a fit-for-purpose format. This occurs during a phase in the generation of ARD called data cleaning (see Section 2.7) and quality assurance (see Section 2.8).

## 2.7 Data Cleaning

The generation of consistent data sets and using a standardised format is a requirement for the generation of ARD. During the process of data cleaning, inconsistencies, and errors are removed [40]. These include missing values, typographical errors, mixed formats, and replicated entries. The removal of these errors increases the overall quality of the data. Through this cleaning process, these inconsistencies and errors can be inspected, removed, or replaced in a consistent, replicable, and justifiable manner for each unique data set. [31], [32]. Data cleaning can be conducted by using open-source software that is widely available and distributed [41]. This provides a fast, reliable, robust, and adaptable way to prepare outputs in a standardised format [42]. Examples of open-source programming languages with libraries widely used for data cleaning: are R, Python, and Structured Query Language [43].

## 2.8 Quality Assurance

Quality assurance is the process of detecting and resolving anomalies in data. This process should be considered when creating DMP and can be conducted simultaneously with data cleaning processes (see Figure 1) [10], [44]. A DMP ensures that the procedures for quality assurance have been established from the onset of a project (see Section 2.4 Data Management and Appendix 3). Quality assurance is a crucial step to improving and guaranteeing the overall quality of the data. A quality assurance process can ensure that ARD are accurate, complete, consistent, relevant and maintain their longevity. Quality assurance is used to identify any missed or unidentified errors in the data cleaning process before outputs are finalised. After the ARD cleaning has been initially complied into a common format, several supplementary quality assurance processes are useful for ARD to be prepared for deployment: data alignment, uncertainty, and privacy preservation.

Data alignment occurs so that the ARD is converted to align across different characteristics so that datasets can be compared with one another. This process typically occurs after the data is cleaned and compiled into a standardised format. In the case of spatial data, data alignment might entail the conversion of geographic or time series data to a standardised areal or temporal scale to facilitate visualisation and comparison. For example, the Australian Bureau of Statistics (ABS) releases regional boundaries that describe population density and communities known as the Australian Statistical Geographical Standard (ASGS) [45]. These regional boundaries are updated every five years and are represented in a nested hierarchical structure of regions. The ASGS is comprised of several different geographical regions which are summarised in Appendix 4. As these boundaries are updated semi-frequently, data needs to be compiled into a standardised format across a standardised areal and temporal scale. This is achieved through correspondence. ABS also releases machine-readable correspondence files for the mathematical conversion between most geographical regions [46].

Uncertainty is often attributed to collection processes (e.g., variance in sampling techniques), variance (e.g., conceptual difference across projects) or multimodality (e.g., data being noisy, inconsistent or incomplete) (see Section 2.6) [2], [38], [39]. This can impact the effectiveness and accuracy of results leading to negative effects on the overall analytics and decisions-making processes [37], [38]. If uncertainty is present in data it can also limit its use [2], [38]. Controlling for uncertainty within the quality assurance process will improve the effectiveness and accuracy of results [37], [38]. Uncertainty should be managed and removed, where possible, during the data cleaning and quality assurance process. If uncertainty cannot be removed it should be acknowledged and documented.

When conducting research using sensitive or identifying information, particularly involving human participants, ARD processes should ensure that adequate steps are taken to provide privacy and protection during the research process. Privacy preservation is the process of removing identifying attributes that are associated with participants [47]. These attributes can include the degree to which an individual is directly identified, typically through an individual's characteristics demonstrated in the data. Risk of re-identification can be minimised through three commonly used techniques: suppression, pseudonymisation, and randomisation. The aim of using these techniques is to achieve an acceptably low risk of re-identification before the data are released to users [48]. The first approach, suppression, removes or suppresses values at a standardised threshold across all records [47], [48], [49]. For example, suppressing all cells with a count less than five is a common threshold for suppression [48], [49]. Pseudonymisation is the implementation of techniques to replace unique identifiers with pseudonyms. These pseudonyms cannot be associated with the data without adequate information (e.g., encryption key) regarding the encryption methods [50]. These can be randomly generated and are used to replace identifiers including names, government issues identifiers or addresses [47]. The final technique, randomisation, adds random values that are not predictable to an existing dataset.

## 2.9 Metadata

An umbrella term for the information used to describe, manage, and relate to the collection of data is referred to as metadata. This information can be generated through automatic or manual processes where a machine-readable output is preferred [51]. Metadata is collected to increase the visibility and reusability of the ARD collected. This visibility and reusability is also considered for Metadata when creating a DMP for a given project. This allows for the metadata creation to be guided by clear and established aims for its creation (see section 2.4 Data Management and Appendix 3).

Metadata principles have been devised to guide the promotion and distribution of data on a broader scale [8], [51]. For example, the Australian Data Research Commons (ARDC) has devised principles for creating standardised metadata. These principles are made up of four components: findable, accessible, interoperable, and reusable (FAIR) data principles [51]. These principles created by the Australian Data Research Commons (ARDC) can be used to maximise data sharing capabilities as well as data use and reuse [8], [51]. ARD can be made findable by providing local and international users with a unique or persistent identifier for each dataset [52]. This allows data to be findable by both humans and computers [51]. ARD can be made accessible through the use of a standardised communication protocol even if the data is not open or available for public use. This allows for there to be clarity and transparency around the access, reuse, and authentication processes required for access to ARD [51], [52]. Interoperable ARD is achieved by providing outputs in a standardised approach. This includes the format, and vocabulary in the ARD and metadata so there can be comparisons between other data and metadata. ARD is established as reusable by providing accurate information to users about licencing agreements and detailed provenance information.

## 2.10  Data Dictionary

Individual data variables are distinct structures and in order to understand their meaning additional information is required [53]. A data dictionary is a central repository of information about variables present in a dataset and their meaning. This information includes the name, origin of the data, data format, usage, cleaning processes, quality assurance processes, and data management tasks [12], [53]. Information present in data dictionaries should be developed and presented in a standardised format. Data dictionaries should also developed in conjunction with metadata, to ensure the interoperability and reusability of ARD [52], [53]. In comparison to metadata (see Section 2.9), data dictionaries provide information on the individual level of data indicators, providing users with supplementary information about the data processes, data sources, and necessary contextual information about the data and support clarity for further research or analytics using a given collection of ARD. Data dictionaries are also useful for the input of new data or updating existing data sets [53]. Additionally, this allows for findings to be conveyed to a wider audience by increasing interpretability and reducing misperceptions [12], [53].

## 3. GUIDELINES IN THE CONTEXT OF A DIGITAL ATLAS

In this section, the proposed guidelines are considered in the context of ARD for a digital atlas. These guidelines were implemented for the Australian Child and Youth Wellbeing Atlas (ACYWA) to map data for children and young people's health and wellbeing across Australia [16]. Data relating to children and young people aged 0-24 years were collated for ACYWA. These data were collected from 2006 to 2022 and were aggregated by gender, age, and geographical region for reasons of privacy. Decisions based on the implementation of the ARD guidelines have been noted in the subsequent sections.

### 3.1 Ethics

In alignment with the proposed guidelines in this paper, ethics approval was gained for ACYWA. This supported the research project creation and design as it abided by the legal and ethical requirements for conducting research in Australia [8], [17]. Furthermore, ACYWA was upheld to high-quality criteria that maintained the security of participant's personal information for the duration of the research process [18]. Appendix 1 outlines key parts of Australia's National Statement, including aspects of the National Statement related to maintaining the security of children and young people. Ethical approval was achieved by the ACYWA development team submitting a Human Research Ethics Application (HREA) through the Human Research Ethics Committee at the Queensland University of Technology. Subsequently, ethics approval was gained for the use of aggregate de-identified data for the purposes of mapping ARD at different geographical levels across Australia using a spatial platform (digital atlas). The HREA submitted for the current project was created and approved in accordance with the National Statement. This specific criteria was chosen as the research involved Australian Universities using data for human participants (Australian children and young people aged 0-24 years old). Consent was not sought for the participants in this project since all data were either publicly available or accessed via request (permission was granted for access and subsequent public release). These data were then de-identified and aggregated at small area levels. These processes were approved by the Human Research Ethics Committee at the Queensland University of Technology.

**3.2 Project Documentation**

A free cross-institutional cloud-based storage system is currently being used for ACYWA data. Project documentation has ensured that there has been effective version control and modification of records for the duration of the research project [21]. Project documentation was established from the onset of research so that scope, research team, risk, communications, liabilities, processes, decisions, data, and other relevant materials for the project were meticulously recorded. This documentation was recorded and shared with and by the core ACYWA development team. These documents are crucial for rare scenarios where handovers of responsibilities may not be feasible, so that researchers will have step-by-step guides from previous researchers through the processes of the past, ongoing, and upcoming work. Project documentation also provided supplementary information during the handover process. This ensures that researchers, current, returning, and future, have sufficient information to maintain this national data asset efficiently and successfully [8], [21]. At current, ACYWA has been designed as an enduring data asset with no foreseeable end date, where updates to security, ethics and data protection methods will be made as required. As the project is currently ongoing, destruction methods have not been implemented. Considerations need to be made by future researchers on ACYWA about the destruction and deletion of project files once the project has been completed. This approach has been approved in the Human Research Ethics Application.

### 3.3 Data Governance

Data governance was employed for ACYWA to safeguard the personal information of participants used in this research. In alignment with the proposed guidelines, the Australian Code for the Responsible Conduct of Research was used to create strategies to protect data from misuse and mishandling as well as adopt safe strategies for collecting, sharing and use of the data [25], [27]. The eight principles honesty, rigour, transparency, fairness, respect, recognition, accountability, and promotion were considered and followed for this research [31]. The approach developed by the ACYWA team was approved in the ethics process.

### 3.4 Data Management

A Data Management Plan (DMP) was completed in alignment with institutional protocols for ACYWA. This DMP underwent an approval process through ethics and was subsequently approved [22], [23]. A summary is available in Appendix 3. As ACYWA is currently ongoing, the DMP is scheduled to be revisited annually [33].

### 3.5 Data Storage

As previously stated, ACYWA utilises a free cloud-based storage system that allows cross-institutional access [35]. The cloud-based storage adopted was OneDrive as it allowed for secure access, editing, and organisation of project documentation and data across multiple institutions [54]. OneDrive was flexible, adjusted to advances in technology, and supported large amounts of data ingestion. This storage system has been reliably accessed by researchers on the project across institutions for data access, upload and project documentation [35], [54]. This has ensured reliable access to shared resources, minimised the risk of loss of data, the safety of data and research information, and that participants' information is secure [8].

### 3.6 Data Discovery and Collection

ACYWA's raw data was collected by first discovering available data resources that aligned with the project participant's characteristics. These characteristics include Australian children and young people aged 0-24 years old. These data were collected from 2006 to 2022 (children and young people born in Australia between 1982-2022) across the available spatial and temporal aggregations. This data was also aggregated by age and sex when available. De-identified data was collected for this project where further de-identification techniques were applied to keep the participant's confidentiality (see sections 2.8 and 3.8). This was achieved by devising a set of potential resources that aligned with the evidenced-based framework developed by the Australian Research Alliance for Children and Youth's (ARACY) The Nest. This framework aims to inform health and wellbeing metrics in order for children and young people to thrive and reach their full potential [55], [56]. The data that was identified as aligning with the desired participant characteristics was then assigned one of six key domains with The Nest: healthy; material basics; valued, loved, and safe; learning; participating; and positive sense of identity and culture. This provided a validated approach in creating, organising, and managing from the onset of the project [8].

These data were then collected from Australian data custodians, and were obtained through custom requests, public access, and customised data downloads (e.g., TableBuilder [57]). The custom requests through data custodians were often subject to significant wait times (the maximum wait for this project was roughly twelve months, from submission of request to data received). Data formatting, requirements, and costs were also discussed at length and these decisions were documented throughout the ACYWA research process. The public data was accessed through resources made available online and the ACYWA development team recorded the stages of the collection process to avoid double handling (e.g., downloading data more than once). This process was followed for all customised data sets to create uniform data sets using the same specifications. These data were downloaded from resources such as TableBuilder, and other government and non-government websites (e.g., Australian Institute of Health and Welfare, Australian Bureau of Statistics, and Australian Government) and were downloaded using a standardised set of specifications.

Documentation surrounding the ARD was collected for the duration of the project. This was beneficial as multiple development team members were working on ACYWA and this allowed them to have sufficient information to conduct research efficiently and successfully [8], [21]. The resulting documentation was compiled into standardised formats and was released alongside the data. This supports the guidelines proposed in this paper, in both the initial phases of research and throughout the entire research project [8].

**3.7 Data Cleaning**

Standardised cleaning processes were used to compile the data for ACYWA into a standardised format. Data was cleaned using open-source software R and RStudio as it provided specialised packages for data cleaning, visualisation, and data analysis [41]. This programming language was chosen as it most closely aligned with researcher's skillset and the ACYWA atlas aims and methodologies. The data was compiled into a standardised form for the ingest into a custom-built platform. This data format is outlined in Table 2. This required adherence to strict regulation of data standards to properly ingest data into the custom-built atlas. Examples of packages used for data cleaning and management include, dplyr, readxlsx, and tidyr [58], [59], [60]. The files for cleaning processes were stored on a cloud-based repository for version control of code and error detection. The resulting data was high-quality ARD that were as free as possible from missing values, typographical errors or replicated entries. Data for ACYWA was compiled in a standardised format that also supported ingest into a custom-built spatial platform. This process improved the quality of the data removed and identified inconsistencies and errors [40]. To date, it is approximated that the current version of ACYWA contains an estimated 430 individual data items which were collected from over 30 different data sources [16]. This data was collected using the ASGS at Statistical Area Two, Statistical Area Three, Statistical Area Four, State and Territory, Australia (National), and Local Government Area (see Appendix 4).

Table 2.

Standardised Data Format used for Geographical Data in ACYWA

| Format | Geography | Filter | Filter | Filter | Data |
|--------|-----------|--------|--------|--------|------|
| Example | SA3CODE_16 | CALENDAR_YEAR | AGE_GROUP | SEX | 0 |

**3.8 Quality Assurance**

Quality assurance processes were used to check ARD quality and data cleaning processes concurrently for ACYWA. If any inconsistencies were found, the cleaning process was revised, and then the quality assurance was reapplied. This process was continued until the researchers were certain that the data was of high-quality, accurate, complete, consistent, and relevant for the long term ingest into the digital atlas [10]. This process was implemented for all data sets in ACYWA. This process was recorded through reproducible documents for reporting and documenting the quality assurance in Rmarkdown [61]. The supplementary quality assurance steps (data alignment, uncertainty, and privacy preservation) were also carried out by the ACYWA development team. Data custodians were also given the opportunity to review the data after ARD processes were applied. Custodian approval was sought on the final ARD output, and any necessary changes were made as per the request of data custodians.

Data for ACYWA was received at several different iterations of the Australian Statistical Geographical Standard (ASGS) boundaries (ASGS2006, ASGS2011, ASGS2016, and ASGS2021) due to the constraints of data availability [62]. As such, correspondence methods, and the standardisation of areal and temporal scales for data comparison, were needed to compile the datasets in a standardised format before the final release onto the digital atlas. ASGS2016 was chosen as the standardised regional boundary as it was the most reliably available at the time of collection. All other regional boundaries (ASGS2006, ASGS2011, and ASGS2021) were converted to ASGS2016 for the standardised comparison of data [1], [3], [4], [5], [6]. Correspondence was carried out through two different techniques: forward correspondence and backward correspondence. Forward correspondence and backward correspondence are explained in more detail in Appendix 5. Additionally, during these correspondence processes, uncertainty was created. This uncertainty is discussed below.

Uncertainty was present at two phases of the data creation process: uncertainty that was already present at the time of collection and uncertainty that was created during ACYWA's alignment processes. Uncertainty present at the time of collection is often due to noisy, inconsistent, or incomplete data (see Section 2.8) [38]. Certain aspects of ACYWA's cleaning and quality assurance processes (e.g., data alignment and correspondence) resulted in uncertainty, as a lead-on effect of the original noisiness, inconsistencies, and incompleteness in the original data. As such, some of the ARD could not be interpreted with absolute certainty in the results and were subsequently removed. This was achieved by first identifying all types of uncertainty and assigning a standardised numerical value depending on how it would impact the accuracy and effectiveness of results [38], [39]. These numerical values were as follows, zero indicated low uncertainty, one indicated medium uncertainty, and two indicated high uncertainty. High levels of uncertainty were then removed in all datasets. Low and medium uncertainty were deemed not to impact the results enough to be removed. If any uncertainty was present in the data, the level was recorded in the data dictionary for ACYWA on an individual data indicator level.

Data for ACYWA was safeguarded through de-identification and suppression techniques. First, the data was provided and downloaded as aggregate summarises (counts, percentages, rates, etc) for each specified geospatial area. This meant individual information was held by the data custodians and that the ACYWA development team did not have access to participants' personal information at any stage during the research process. Second, suppression techniques were applied by both the data custodians and the ACYWA development team. Suppression was applied across all data sets wherever cell counts of less than five were present [48], [49]. These approaches were documented, included in the DMP, and submitted to ethics for subsequent approval. These techniques ensured that the data risk of misuse and mishandling of ACYWA data was minimised through the protection of participants and their personal information [25], [27], [48], [49].

## 3.9 Metadata

Data used in ACYWA has been made accessible with increased visibility and reusability by implementing the FAIR principles created by ARDC [8], [51]. The principles adopted were devised to create metadata specifically for geographical and spatial data that aided in the data sharing capabilities for use in a digital atlas [8], [51]. ACYWA metadata was created using both manual and automatic processes [51]. An automatic process through Posit's Rmarkdown was used to produce a summary table with the necessary information to be included in the metadata. This information was then manually transferred to a standardised metadata document, alongside other information that was manually inputted that could not be sourced from Rmarkdown. A standardised output was then produced for individual datasets. ACYWA's metadata aligned with the four key aspects of ARDC's FAIR principles. This was achieved by utilising a customised version of UWA-PURE that includes the core mandatory elements of additionally the metadata standard ISO19115. For each data set a metadata document was created and as required, was sent for approval and feedback to the appropriate data custodians. These documents were then compiled into a single document which is freely available online and accessible via the ACYWA website. The key aspects of the ACYWA's metadata that aligned with the principles have been outlined in Table 3.

**Table 3.**

Summary of the Key Metadata Elements

| ARDC FAIR Principles | Examples of ACYWA elements that align |
|---|---|
| Findable | - Title (e.g., name of original source)<br>- Digital Object Identifier<br>- Metadata reference |
| Accessible | - Legal and ethical requirements for data use<br>- Access rights and visibility for data sharing |
| Interoperable | - Metadata has been compiled in a standardised approach using broadly applicable language and information |
| Reusable | - Licence<br>- Geographical coverage<br>- Temporal coverage<br>- Fields of research<br>- Socio-economic objectives |

**3.10 Data Dictionary**

The data dictionary for ACYWA captures the necessary information that provides insight to users into the creation, usage, and format of the data [12]. The data dictionary for each variable was created manually by the researchers on ACYWA and has undergone a quality assurance process. These items will be provided to two different end users: users of the data and members of the ACYWA development team. These items have been summarised in Table 3, where the data dictionary items for users are planned for future release. The information provided to researchers is stored securely, to maintain the integrity of the project, minimise reputational risk and does not risk the identities of the participants in the project [8]. A similar process was followed in the creation of metadata, where Posit's Rmarkdown created a summary output with the key information and the ACYWA development team manually filled in the entries for the data dictionary. The entries not covered by the Rmarkdown output are then entered manually be the researchers. These entries then undergo a quality assurance process. This allows for the data to have maximised visibility, reach a wider audience and avoid any misperceptions when the data is being interpreted [12], [53].

The key items that have been included in the data dictionary for ACYWA have been summarised in Table 4.

**Table 4.**

Key elements in the ACYWA data Dictionary for users and researchers

| Data Dictionary Users | Examples of Data Dictionary Elements in ACYWA |
|---|---|
| Users of Data (published) | - Data variable name<br>- Data variable Definition<br>- Variable Type (count percentage, etc)<br>- Data Source<br>- Application of temporal correspondence<br>- Presence of uncertainty |

| Researcher-Only (not published) | <ul><li>Links for<ul><li>Cleaning Code</li><li>Data files</li><li>Project documentation</li></ul></li></ul> |
|---|---|

## 4. CONCLUSION

The current paper aimed to contextualise current approaches into a new set of guidelines that can be used to generate ARD. A total of ten steps were developed that can be used to produce ARD: Ethics, Project Documentation, Data Governance, Data Management, Data Storage, Data Discovery and Collection, Data Cleaning, Quality Assurance, Metadata and Data Dictionary. These guidelines have effectively combined previous literature to provide a comprehensive approach to ARD creation. Together, these principles form the first focused framework for producing ARD. The guidelines were also successfully contextualised to produce ARD for the digital platform ACYWA.

ARD processes are crucial for facilitating the comparison, inspection, and display of different geospatial and temporal data [15]. The paper showcased an example of effective the implementation of this guideline in the context of a digital atlas (ACYWA). Thus, achieving the aim of compiling current literature in a comprehensive guide for the creation of ARD. The resulting output has created a data asset that is free and publicly available to a range of users, and stakeholders, for general public use which can be used to inform children and young people's health and wellbeing metrics in Australia [16]. This digital atlas provides users with high-quality ARD on multiple different spatial and temporal levels across Australia that was not previously available. The guidelines also provided the development team with meaningful insights into the overall creation of the project. This includes the generation of supplementary materials such as project documentation, a data management plan, metadata, and a data dictionary. Overall, the culmination of the output of high-quality ARD and associated documentation by adhering to stricter generation processes promotes more trust in the data and the resulting product (i.e. a digital atlas).

The final product for the ACYWA is online, free, and publicly available for users, but establishment, collection, cleaning had some associated costs. This should be factored into future projects where data sources will need to be chosen subject to the availability of funding. For example, should data be collected from only free and public resources only or is there availability for paid resources as well. Data custodians for ACYWA also had the final decision on the inclusion or exclusion of data in the digital platform. Data could meet all ARD requirements, but might not meet a data custodian's individual, internal approval processes. As such, a few custodians did not allow the ARD to be shared further than the ACYWA development team and could not be shared publicly. This decision could only be made after all ARD processes had been applied and the final product could be reviewed by data custodians. Albeit, this was a rare occurrence, where an estimated 430 individual data items were included in ACYWA. Nonetheless, the exclusion of ARD outputs due to custodian review processes should still be considered as a possibility for future research. Additionally, the standardised data format was created with the specifications required for geographical data and ingested into a digital atlas. This standardised format can be contextualised to other digital atlases and geographical data that require the same specifications. The use of this standardised approach allows the data to align with FAIR data principles, especially if the format is adopted by other projects. Future work could explore the use of this standardised approach to different types of data.

Strategies could be used to evaluate the compliance of researchers when implementing these guidelines in future studies. This research could also involve exploring the efficacy of the different steps and whether all steps are needed to produce the same high-quality ARD output. For example, if either a data management plan or quality assurance are excluded will the same output be produced and is the output of high-quality. The reordering of steps could also be explored. Different research fields may require a varied order of steps. This can be explored in future case studies if implementation of the steps in a different order yields successful production of ARD. Future work could also be undertaken to modify the ARD generation steps to focus on different core principles depending on the field the research is being conducted. For example, satellite data will require different generation steps than data using human participants. More specifically, satellite data may not require ethics, but would likely place a larger focus on data cleaning, quality assurance, and metadata processes where data needs to be aligned temporally, and be made findable, accessible, interoperable, and reusable [1], [3], [14]. The current guidelines provide a framework and a general guide needed for the generation of ARD.

These guidelines can be used to not only to improve the overall quality of ARD through the stages such as discovery and collection, cleaning, as well as quality assurance; but to ensure that the overall project is conducted at a high standard by considering steps such as ethics, project documentation, and data storage. This ensures that the data that was initially limited in its use is made accessible to its desired users [2]. As a result, this allows for the expansion of use through the trust in high-quality ARD outputs by users (e.g., the general public, stakeholders, users, researchers, etc) [16]. These uses can include new policy decisions, analytics, or general research. This also promotes further discussion in the scientific community and broader fields of research for the development of ARD through comprehensive guidelines.

The new guidelines build on and synthesise existing literature and have been showcased using a digital atlas. These guidelines allow for further consideration of processes underpinning the creation of ARD and steps, allowing avenues for innovation, novel, and future research. Overall, these guidelines provide a systematic approach to ARD creation that incorporates insights from existing literature and leveraging digital atlas technology that was not previously available.

**AUTHOR CONTRIBUTIONS**

Harriette Phillips (harriette.phillips@hdr.qut.edu.au), Owen Forbes, Aiden Price, Claire Boulange, and Marketa Reeves were involved in the data collection processes for the Australian Child and Youth Wellbeing Atlas, which led to the development of this paper. Harriette Phillips wrote the manuscript. Kerrie Mengersen and Aiden Price provided supervision of the original draft, revision and proofreading of the resulting manuscript. Owen Forbes, Claire Boulange, Marketa Reeves, Rebecca Glauert provided revisions and proofreading to the resulting manuscript. All authors listed have provided meaningful and valuable contributions to the manuscript.

**ACKNOWLEDGEMENTS**

This work was supported by the Queensland University of Technology's Centre for Data Science and Research Training Program Stipend Scholarship. The work was also supported by the Australian Child and Youth and Wellbeing Atlas.

**DATA AVAILABILITY STATEMENT**

The ARD that has been produced is free and publicly available through the ACYWA platform. The digital platform can be found here: https://australianchildatlas.com/.

# APPENDIX

## Appendix 1

Overview of the General Consideration and Considerations for ACYWA for the National Statement of Ethical Conduct of Human Research [20].

| Aspect of National Statement Overview | |
|---|---|
| Merit and Integrity in Research | <ul><li>Research outputs can be justified by research benefits.</li><li>Research conducted, developed, or designed using appropriate methodology which align with the aims of the given research proposal.</li><li>Research is based on a thorough study of current literature and previous studies. This includes research in quick response to unforeseen situations and possible novel research.</li><li>Research aims of the research, how it is carried out, and the results should ensure that conducted in such a manner that respects the participants of the research.</li><li>Facilities where the research is conducted is appropriate for the given topic.</li><li>Qualification or competence of researchers are sufficient to supervise or conduct research.</li><li>When conducting research, the researchers should be researching for knowledge and understand of the topic, follow the principles of research conduct, disseminate results whether favourable or unfavourable.</li></ul> |
| | - Once a peer review of the research has been completed, research merit is no longer dictated by those completing the ethical review. |
| Justice | - The results of the research have been accurately and fairly described.<br>- The scope, research objectives, and inclusion and exclusion criteria have been considered.<br>- Participants are recruited fairly, are not burdened, or exploited in the research process.<br>- Research outputs are disseminated in a timely and clear manner, and any benefits are made available to participants |
| Beneficence | - Any risk of harm or discomfort to participants must be outweighed by the benefits of the research. The research must be likely to show benefit to either the participants, their community, or both.<br>- Researchers should design research to minimise risk and discomfort, as well as making any potential risks or benefits known to participants, whilst keeping the welfare of participants in mind. |
| Respect | - The confidentiality, cultural sensitivities, and privacy of participants and their communities should be respected by researchers and institutions when conducting research. If necessary, any specific agreements should between participants or communities should be fulfilled during research.<br>- All research should take into consideration the autonomy of humans, and their ability to make decisions on their own. If participants have a diminished capacity for decisions making, researchers should empower participants where possible to make decisions and provide them with the necessary protections. |

| Risk and Benefits | <ul><li>Any risks, harm, discomfort, or inconvenience (risk) should be identified, assessed for likeliness to occur, identified who the risk is likely to affect, how risk will be minimised, potential benefits, and who the benefits are likely to affect.</li><li>The risks and severity of risk should be reviewed based on available evidence.</li><li>Research can only be deemed "low risk" if the risk identified in the research is "discomfort". Research is identified as "negligible risk" if the risk of harm or discomfort is not foreseeable.</li></ul> |
|---|---|
| General Requirements for Consent | <ul><li>Research participants should participate in research voluntarily. Participants should also be provided with sufficient information about the research project (e.g., methods, risks, outputs, demands, or potential benefits) and the implications of the research to make an informed decision about their involvement.</li><li>Participants must be provided with information that is presented in a manner that is suitable for all.</li><li>Consent needs to be gained for the inclusion in research projects where participants gain an understanding of their involvement in the research process. Consent can be gained in the following ways:<ul><li>Verbally</li><li>Written</li><li>By other means (e.g., Survey)</li></ul></li></ul> |
| ACYWA Specific Aspect of the National Statement Overview | |
| Children and Young People | - Consent should be gauged for the inclusion of children and young people participation in research.<br>- Researchers should describe how children and young people will be described the research and how it will be relayed to them at a level they can comprehend.<br>- If consent cannot be obtained, inclusion in the research must be justified by advancement in knowledge or children and young people's involvement are indispensable to the research. Research should be conducted in the best interested of children and young people.<br>- Research should be conducted in manner that provides an emotional and psychological security, safety, and wellbeing for children and young people. |

## Appendix 2

Outlines of the key eight principles in the Australian Code for the Responsible Conduct of Research

| Principle | Overview |
|---|---|
| Honesty | - Truthful and accurate information presented when preposing, conducting, and reporting research |
| Rigour | - Robust methodology used, attention to detail to avoid biases which are otherwise acknowledged |
| Transparency | - Research methodology, data, and findings are communicated and shared openly |
| | - Research is communicated responsibly and accurately<br>- Conflicts of interests are managed or disclosed |
| Fairness | - Researchers and others involved in the research should be treated with respect and fairly<br>- Works of others should be cited and referenced<br>- Appropriate credit should be given to contributors to the research |
| Respect | - Conducting research for minority and vulnerable people should be given appropriate considerations, ensuring that participants and communities affects by research are care for and respected |
| Recognition | - Conducting research with Aboriginal and Torres Strait Islander Peoples should be engaged in the research which affects or is significance to them<br>    ○ Aboriginal and Torres Strait Islander Peoples connection to land, diversity, heritage, knowledge, and cultural property is respected, valued, and recognised<br>    ○ Prior to undertaking research, Aboriginal and Torres Strait Islander people are engaged with so that they may freely make decisions about their involvement<br>- The outcomes of the research is reported to the Aboriginal and Torres Strait Islander people who engaged with the research |
| Accountability | - Guidelines, policies, and legislations should be abided by<br>- Good stewardship used when conducting research with public resources<br>- Consequences and outcomes considered prior to communication |

| Promotion | • Responsible conduct of research is fostered and promoted |

**Appendix 3**

Data Management Plan topics for consideration

| Key processes | Overview |
|---|---|
| Data Discovery | • What is the expected type of data that is going to be collected for the project? |
| Data Description | • Provide a description of the data being collected for the data (i.e., what are the specifications for the data, who are the participants, etc). |
| Data Collection | • Who will be in charge of collected the data?<br>• Where will the data collection occur?<br>• Are there any equipment requirements for the collection of the data?<br>• What are the expected characteristics of the data?<br>• What is the start date of the data collection?<br>• What is the end date of the data collection?<br>• What is the expected volume of data? |
| Data Organisation | • How will the data be stored once it is received?<br>• How will this be catalogued and documented?<br>• Are the quality control measure in place? |
| Data Preservation Procedures | • Are the data preservation procedures beyond the scope of the project? |
| Data Storage | • Where will the data be stored? <br> • Does this storage account for data loss (see data loss procedures)? |
| Data Volume | • What is the expected volume of data for the research project? <br> • Can the chosen data storage handle the expected volume of data? |
| Data Loss Procedures | • Are there multiple storage options to account and prevent data loss? <br> • If data loss occurs, are there appropriate measure in place to negate the loss of data? |
| Data Privacy and Confidentiality | • Participants privacy and confidentiality been considered? <br> • Is there no/ low risk to participants? <br> • How have risks been minimised? |
| Data Ownership | • Which organisation or individual has ownership of the data collected? |
| Quality Assurance | • What quality procedures are in place? |
| Documentation | • Where are decisions made regarding the data collection processes stored? <br> • Who is documenting decisions? <br> • Are decisions about the data being documented? <br> • What decisions should be documented? |
| Dissemination | • Is the data being disseminated? <br> • If so, how is it being disseminated? <br> • Have legal, licencing, and ethical requirements been considered? |
| Metadata | • Has metadata principles been completed to make the data more accessible? |

**Appendix 4**

An overview of the structure of regions in the Australian Statistical Geographical Standard (ASGS) which was created by the Australian Bureau of Statistics *[62]*.

| Australian Statistical Geographical Standard | |
|---|---|
| Mesh blocks | • Mesh Blocks are the smallest defined geographical area and are used as define the larger regions in the ASGS<br>• Mesh blocks typically contain 30-60 dwellings |
| Statistical Area Level 1 (SA1) | • Used for the description of census data<br>• SA1 typically contains 200-800 people |
| Statistical Area Level 2 (SA2) | • SA2s are used to represent communities, and how they interact with one another socially and economically<br>• SA2s typically contain 3, 000 to 25, 000 people |
| Statistical Area Level 3 (SA3) | • SA3s are designed for the outputs of regional data<br>• SA3s typically contain 30, 000 to 130, 000 people |
| Statistical Area Level 4 (SA4) | • SA4s function as defining key area, such as capital cities, representing regional data and the labour markets<br>• They typically contain over 100, 000 people |
| State and Territory (STE) | • The legally defined, cartographic representation of the state and territory boundaries |
| Australia | • A representation of the entirety of Australia |
| Indigenous Structure | • Used for the analysis and publication of statistics for the Aboriginal and Torres Strait Islander population in Australia |

| Remoteness Structure | • Five classes of remoteness (major cities, inner regions, outer regional, remote, and very remote) based on relative access to services |
|---|---|
| Non-ABS Structures | • Structures that are not defined by the ABS<br>    o Local Government Areas (LGA)<br>    o Destination Zones<br>    o Australian Drainage Regions<br>    o Tourism Regions<br>    o Postal Areas<br>    o State Electoral Divisions<br>    o Suburbs and Localities (formerly State Suburbs) |

**Appendix 5**

Backward correspondence is the process of estimating count data for ASGS2016 regions using count data from ASGS2021 regions. Forward correspondence is the process of estimating count data for ASGS2016 using count data from ASGS2006 and ASGS2011 regions. That is, this approach achieves standardisation of the data backward through the temporal scales across the same areal levels. Consider a region, $A$, which has been split into two regions, $B$ and $C$, through known ratios, $a$ and $b$, respectively (see Figure 2a). Then forward correspondence is calculated as follows for regions $B$ and $C$.

$$B = a \times A$$

$$C = b \times A$$

with $a + b = 1$ such that region $A$ is distributed entirely to ASGS2021 regions $B$ and $C$. To derive ASGS2016 region counts from ASGS2021 regions, i.e., perform backward correspondence, consider the original distribution of counts in reverse. In Figure 2b, for example,

$$A = a \times B + b \times C,$$

where all of regions $B$ and $C$ combine to form region $A$. i.e., $a = b = 1$. Consider now the more complex example in Figure 3a, where two ASGS2016 regions, $A$ and $B$, are distributed to three ASGS2021 regions, $C$, $D$ and $E$. Region distributions between ASGS years, or forward correspondence, is given by

$$C = a \times A,$$

$$D = b \times A + c \times B,$$

$$E = d \times B.$$

Backward correspondence for this scenario is visualised in Figure 3b, with counts for regions $A$ and $B$ calculated as follows.

$$A = a \times C + b \times D,$$

$$B = c \times D + d \times E.$$

In this process, regions $C$ and $E$ distribute uniquely to regions $A$ and $B$, respectively, and so $a = d = 1$. However, region $D$ in ASGS2021 was formed by two ASGS2016 regions, $A$ and $B$, and so must redistribute its count data to both regions, with $b + c = 1$. In this case, however, we do not know either $b$ or $c$. That is count estimates distributions are not known throughout region $D$ for the purpose of redistribution to regions $A$ and $B$. To provide correspondence measures when this occurs, the above calculations are considered using the forward correspondence ratios in Figure 3a as follows.

$$A = \begin{cases} C, & b < 0.1 \\ Suppressed, & b > 0.1 \end{cases}$$

$$B = \begin{cases} E, & c < 0.1 \\ Suppressed, & c > 0.1 \end{cases}$$

In this way, insignificant contributions are discarded (less than 10%) from any backward correspondence calculation. As a result, the backward correspondence process described above can occasionally introduce uncertainty into data assets. This uncertainty is discussed in sections 2.8 Quality Assurance and 3 Guidelines in the Context of a Digital Atlas.

**Figure 3.**

Visualisation of the Forward Correspondence Process

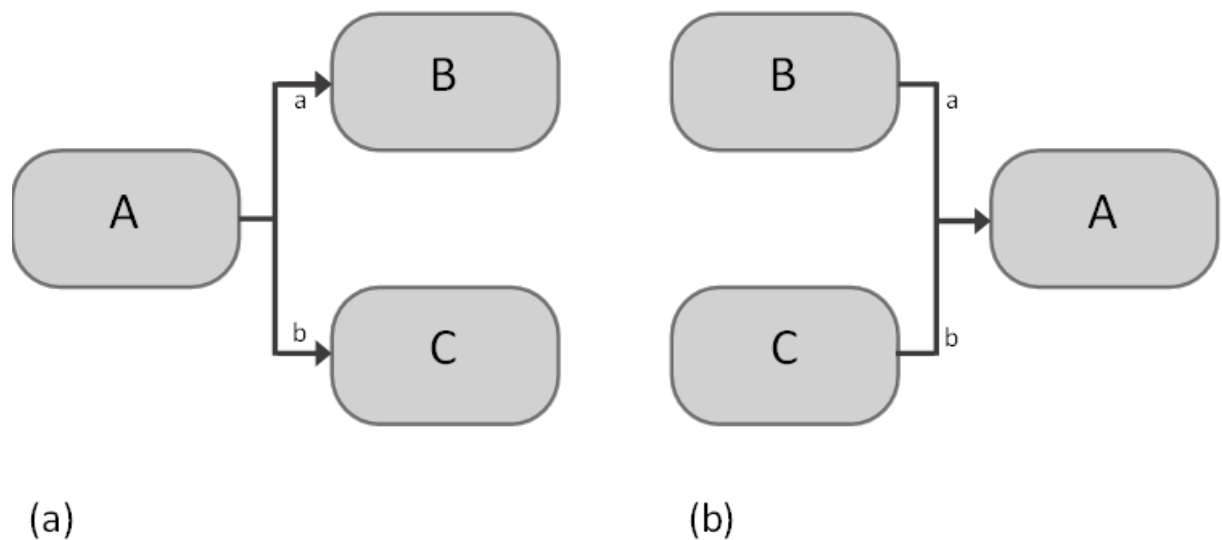

**Figure 4.**

Visualisation of the Backward Correspondence Process

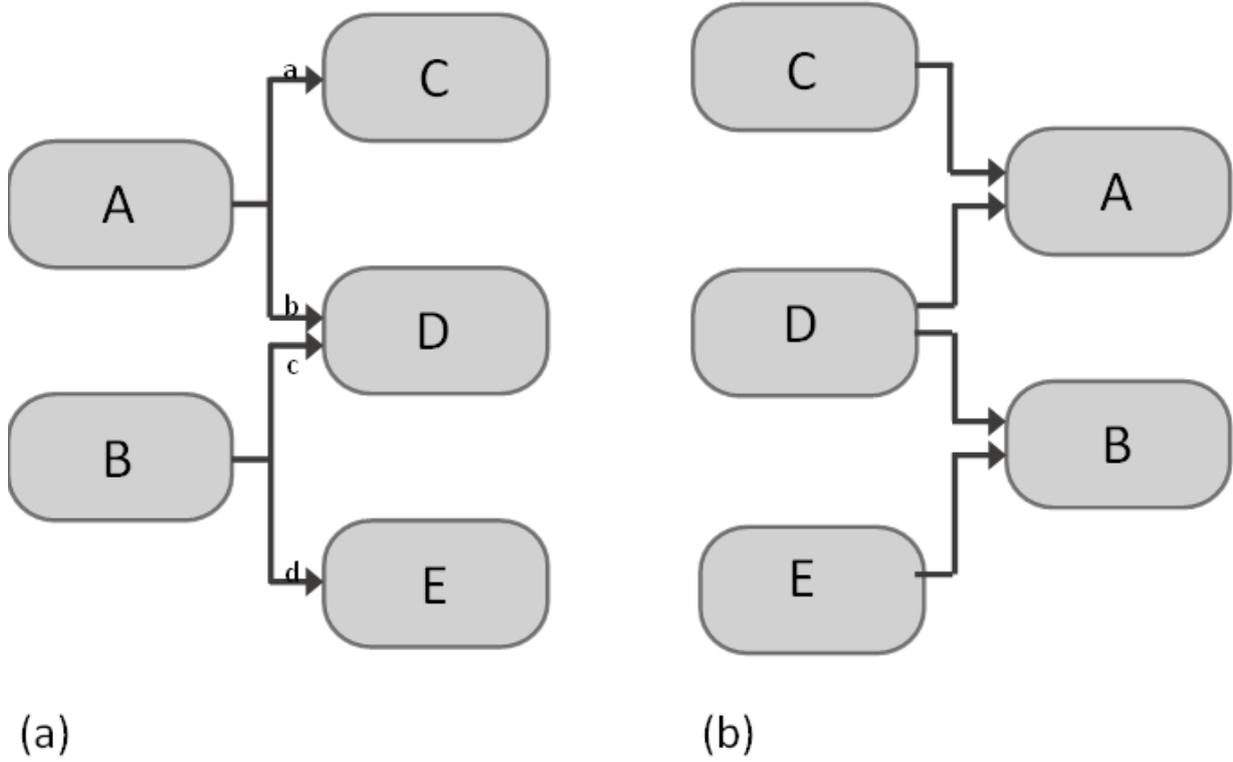